\documentclass[9pt,twocolumn,twoside]{osajnl}

\journal{ol} 

\setboolean{shortarticle}{true}

\title{Optomagnets in nonmagnetic plasmonic nanostructures}

\author[1]{Vage Karakhanyan}
\author[1]{Yannick Lefier}
\author[1]{Cl\'ement Eustache}
\author[1,*]{Thierry Grosjean}

\affil[1]{Optics Department – FEMTO-ST Institute UMR 6174 - Univ. Bourgogne Franche-Comt\'e – CNRS - Besançon, France}

\affil[*]{Corresponding author: thierry.grosjean@univ-fcomte.fr}

\begin{abstract}
Using a  hydrodynamic model of the free electron gas of a 
metal, we theoretically investigate optically-induced DC current loops 
in a plasmonic nanostructure.  Such current loops originate from an 
optical rectification process relying on three electromotive forces, 
one of which arises from an optical spin-orbit interaction. The 
resulting static magnetic field is found to be maximum and dramatically confined at the corners of the plasmonic nanostructure, which reveals the ability of metallic discontinuities to concentrate and tailor 
static magnetic fields on the nanoscale. 
\end{abstract}

\setboolean{displaycopyright}{true}

\begin{document}

\maketitle


Optically-induced magnetism has drawn considerable interest in the past years for its ability to speed up magnetic processes \cite{maccaferri:jap20}. For example, static magnetic fields have been demonstrated to be generated in non-magnetic plasmonic (gold) nanoparticles and nano-apertures \cite{smolyaninov:prb05,Cheng:np20}. Such a phenomenon has been analyzed as the result of the inverse Faraday effect \cite{nadarajah:ox17,hurst:prb18,mondal:prb15}. 

Here, we investigate optically-induced static magnetic fields from a classical hydrodynamic model of the free-electron gas of a metal. Upon illumination with circular polarization, three electromotive forces are identified to produce a tiny DC current loop in a non-magnetic plasmonic nanostructure via an optical rectification process. One of these forces arises from an optical vortex generated within the plasmonic nanostructure by optical spin-orbit interaction \cite{bliokh:sci15,gorodetski:prl13}. The two other forces lead to current distributions localized at the surface of the particle. All of these three forces mediate the transfer from the spin angular momentum (SAM) of the incoming light to an extrinsic orbital angular momentum (OAM) in the free electron gas (linked to a curved trajectory of free electrons). The optically-induced static magnetic field is found to be maximum and to undergo sharp gradients at the corners of the plasmonic nanodisk. This is a consequence of the ability of surface plasmons to be dramatically confined and enhanced at metal discontinuities \cite{pile:apl05}. Surface plasmons at metal discontinuities may thus offer new opportunities to locally concentrate and tailor static magnetic fields on the nanoscale.


The DC component of an optically-induced current density in a plasmonic nanostructure can be deduced from the distribution of an electromagnetic field within the metal via a hydrodynamic model of the free electron gas. The free electron dynamics inside the metal then reads \cite{scalora:pra10,lefier:thesis}:

\begin{equation}\label{eq:hydrobrute}
\frac{\partial \mathbf{v}}{\partial t} + (\mathbf{v} \cdot \mathbf{\nabla}) \mathbf{v} = -  \frac{1}{\tau} \mathbf{v} +  \frac{e}{m} \mathbf{E} + \frac{\mu_0 e}{m} \mathbf{v} \times \mathbf{H}
\end{equation}

where $m$, $n$, $\tau$ and $\mathbf{v}$ are the effective mass, the density, the collision time and velocity of the free electrons, respectively. $\mathbf{E}$ and $\mathbf{H}$ are the electric and magnetic optical fields, respectively. $n$, $\mathbf{v}$, $\mathbf{E}$ and $\mathbf{H}$ are space dependent.

Identifying the current density with $\mathbf{j}=n e \mathbf{v}=\dot{\mathbf{P}}$ ($\mathbf{P}$ is the polarization density vector) and considering small variations of the electron density regarding the equilibrium charge density ($n_0$) in the absence of any applied fields, Eq. \ref{eq:hydrobrute} can be rewritten as \cite{scalora:pra10}:

\begin{equation}\label{eq:eqJ}
\frac{\partial \mathbf{j}}{\partial t} + \frac{\mathbf{j}}{\tau} = \frac{n_0 e^2}{m}\mathbf{E} +\frac{\mu_0 e}{m} \mathbf{j} \times \mathbf{H} -\frac{e}{m}(\mathbf{\nabla}\cdot \mathbf{P}) \mathbf{E} + \frac{1}{n_0 e} \left[ (\mathbf{\nabla}\cdot \mathbf{j})\mathbf{j} + (\mathbf{j}\cdot \mathbf{\nabla} )\mathbf{j} \right]
\end{equation}

In the time-harmonic regime, the solution of Eq. \ref{eq:eqJ} reads 

\begin{equation}\label{eq:solj}
\mathbf{j} = (\mathbf{j}_{\boldsymbol{\omega}} \exp(i\omega t) + c.c)+\mathbf{j_d},
\end{equation}

where $\omega$ is the angular frequency, $t$ is time, $c.c$ is the complex conjugate and $\mathbf{j_d}$ is the continuous drift current induced by optical rectification within the metal. Being of smaller amplitude, the three non-linear terms on the right side of Eq. \ref{eq:eqJ} can be considered as a perturbation of the fourth linear term $\frac{n_0 e^2}{m}\mathbf{E}$ (as $\mathbf{j_d}$ regarding $\mathbf{j_\omega}$). In the zeroth order approximation, Eq. \ref{eq:eqJ} leads to the solution :

\begin{equation}\label{eq:jomega}
\mathbf{j}_{\boldsymbol{\omega}}= \gamma_{\omega} \mathbf{E}_{\boldsymbol{\omega}},
\end{equation}

where $\gamma_\omega = \gamma_0/(1-i \omega \tau)$ and $\gamma_0 = n_e~e_e^2~\tau/m_e$ are the dynamic and static conductivity of the metal, respectively. We have $\mathbf{P}_{\boldsymbol{\omega}}=\mathbf{j}_\omega/(i \omega) \mathbf{E}_{\boldsymbol{\omega}}$. Then, the drift current density $\mathbf{j_d}=\langle \mathbf{j} \rangle$ is obtained by injecting Eq. \ref{eq:jomega} in the time-averaged expression of Eq. \ref{eq:eqJ} at the first-order perturbation approximation, which is relevant to the optical rectification process.

Because Eq. \ref{eq:eqJ} refers to the fundamental principle of the dynamics, the four terms on its right side are homogeneous to electromotive forces. 
Being quadratic regarding the optical fields, three of these forces are relevant to second-order non-linear optical effects including optical rectification. 

The "magnetic force" (Eq. \ref{eq:fmag}) is proportionnal to the Poynting vector of the electromagnetic wave within the metal.   

\begin{equation}\label{eq:fmag}
\mathbf{f_m} = \frac{\mu_0 e}{m}\gamma_\omega~\mathbf{E}_{\boldsymbol{\omega}} \times \mathbf{H}_{\boldsymbol\omega}
\end{equation}

The "quadrupolar force" (Eq. \ref{eq:fquad}) originates from the inhomogeneities of the electron density (characterized by the divergence of the polarization $\mathbf{P}$). The gradient of electron densities being maximum at metal/dielectric interfaces, such a force can be considered as a surface force. 

\begin{equation}\label{eq:fquad}
\mathbf{f_{quad}}= -\frac{e}{m}(\mathbf{\nabla}\cdot \mathbf{P}_{\boldsymbol{\omega}}) \mathbf{E}_{\boldsymbol{\omega}}
\end{equation}

The last "convective" force (Eq. \ref{eq:fconv}) is also dependent on the spatial variations of the polarization $\mathbf{P}$. It is thus also a surface force that is enhanced in nano-patterned metals.    
 
\begin{equation}\label{eq:fconv}
\mathbf{f_{conv}}= -\frac{1}{n_0 e} \left[ (\mathbf{\nabla}\cdot \mathbf{j}_{\boldsymbol{\omega}})\mathbf{j}_{\boldsymbol{\omega}} + (\mathbf{j}_{\boldsymbol{\omega}}\cdot \mathbf{\nabla} )\mathbf{j}_{\boldsymbol{\omega}} \right]
\end{equation}

Therefore, the drift current density $\mathbf{j_d}$ can be written as:

\begin{equation}\label{eq:jdNL}
\mathbf{j_d}=  \tau \left( \langle \mathbf{f_{mag}} \rangle + \langle \mathbf{f_{quad}} \rangle + \langle \mathbf{f_{conv}} \rangle \right)
\end{equation}

where $\langle \mathbf{f} \rangle$ denotes the time averaged component of the force. All of these three forces being of the form $\mathbf{f}=\mathbf{A}~op~\mathbf{B}$, where "`$op$"' is a product-type operator, then we have $\langle \mathbf{f} \rangle = \frac{1}{2} \mathfrak{Re} \left( \mathbf{A}~op~\mathbf{B}^*  \right) $.


The resulting static magnetic field can be predicted from the current density $\mathbf{j}(\mathbf{r'})$ with the Biot et Savart relation: 

\begin{equation}\label{eq:Biot}
\mathbf{B}(\mathbf{r})= \frac{\mu_0}{4 \pi} \iiint_V \frac{\mathbf{j_d}(\mathbf{r'}) \wedge (\mathbf{r} - \mathbf{r'})}{|\mathbf{r} - \mathbf{r'}|^3} ~ d^3r',
\end{equation}

where $\mathbf{r}$ and $\mathbf{r'}$ are space coordinate vectors. $\mathbf{j_d}$ can also lead to the magnetic moment \cite{jackson:book99}:

\begin{equation}\label{eq:momentmag}
\mathbf{m}=\frac{1}{2}\iiint_V \mathbf{r} \wedge \mathbf{j_d}~d^3 r.
\end{equation}

Since $\mathbf{j_d}= e n \mathbf{v_d} = \frac{e}{m} \mathbf{p_d}$, where $\mathbf{p_d}$ is the density of linear momentum of the free electrons,  we have:

\begin{equation}\label{eq:momentangmag}
\mathbf{m}=\frac{e}{2m} \mathbf{L_d}
\end{equation}

where $\mathbf{L_d}$ is the extrinsic orbital angular momentum characterizing the DC current density. It results from the transfer of the spin angular momentum of light to the free electron gas via the three above-described electromotive forces. 

\begin{figure}[hbt!]
\centering
\includegraphics[width=1\linewidth]{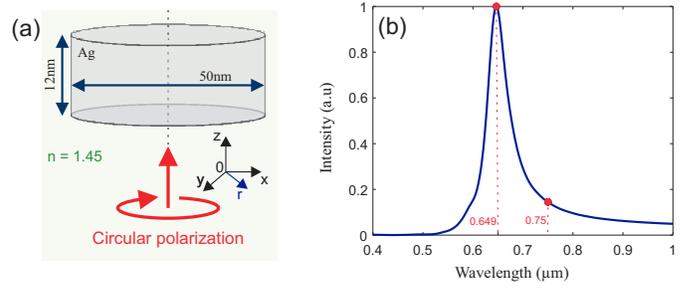}
\caption{(a) Schematics of the optical configuration under study. (b) Spectral response of the plasmonic nanostructure.}\label{fig:cyl}
\end{figure}

We consider a 50 nm-diameter and 12 nm high silver cylinder in a medium of refractive index equal to 1.45 (Fig. \ref{fig:cyl}(a)). The metallic nanostructure is illuminated with a right-handed circularly polarized gaussian beam (beam waist at focus equal to 0.324 $\mu$m) propagating along the axis of symmetry of the nanocylinder. According to our convention, right circular polarization corresponds to electromagnetic fields rotating clockwise when ones look toward the propagation direction of the optical wave. All numerical electromagnetic simulations are realized using the 3D FDTD method. The nonuniform grid resolution varies from 30 nm for portions at the periphery of the simulation to 0.5 nm within and near the nanodisk. 

The spectral response of our nanostructure (Fig. \ref{fig:cyl}) reveals an optical resonance centered at $\lambda$=649 nm, where $\lambda$ is the wavelength. This spectrum is obtained with a Gaussian excitation described by a single temporal pulse. The time-varying electric field is calculated at a single cell located 5 nm away from the side of the nanocylinder. The same simulation is repeated without the nanoparticle. The spectral response is then calculated by Fourier transforming these two results and normalizing the first spectrum (with the cylinder) by the second one (without). 

In continuous wave regime, calculations are conducted in and out of the plasmonic resonance (at wavelengths of 649 nm and 750 nm, respectively). The optically-induced magnetism is represented in a longitudinal plane ($x0z$), where $(0z)$ matches the symmetry axis of the nanocylinder. First, the electromagnetic fields are calculated across the ($x0z$)-plane by FDTD. Then, the distribution of DC current is deduced from the expressions of the electromotive forces given above. To this end, the conductivity $\gamma_\omega$ of the metal at optical frequencies is written as: 

\begin{equation}
\gamma_\omega=i\omega\epsilon_0(\epsilon_\omega-1),
\end{equation}

where $\epsilon_\omega= \epsilon' + i~\epsilon^{''}$ is the complexe relative permittivity of the metal ($i=\sqrt{-1}$) and $\epsilon_0$ is the permittivity of vacuum. $\epsilon_\omega$ is an experimentally accessible optical parameter of materials which merges conductivity from free electrons and absorption from bound electrons (e.g., via electronic inter-band transitions). Deducing $\gamma_\omega$ from $\epsilon_\omega$ thus requires to work in a spectral domain where the free electrons mainly drive the optical properties of the metal (i.e., a spectral domain where $\epsilon_\omega$ is accurately described with a Drude model).	Light absorption would otherwise be artifactually interpreted to contribute to the generation a DC current within the metallic nanostructure. All considered wavelengths in the current paper fulfill the above-mentioned condition for silver
\cite{johnson:prb72,novotny:book}. Finally, the static magnetic field and magnetization are obtained from the distribution of current density using Eqs. \ref{eq:Biot} and \ref{eq:momentmag}, respectively. The magnetic force exerted on a particle of magnetic moment $\mathbf{m}$ reads $\mathbf{F}=\nabla \left(\mathbf{m} \cdot \mathbf{B} \right)$. 

Given the cylindrical symmetry of our system, the DC currents solely arise from the azimuthal component of the three electromative forces. The longitudinal component of the three forces (along (0z)) induce a charge accumulation at the circular interfaces of the particle. When this charge accumulation reaches saturation, the longitudinal current vanishes and a DC voltage appears, a phenomenon described as an optical Hall effect \cite{ashcroft:book}). Given the limited extension of the nanostructure, the radial component of the forces can neither lead to a DC current. 

Being proportional to the Poynting vector, the magnetic force (Eq. \ref{eq:fmag}) develops a DC current loop when the incoming optical waves show a helical phase gradient (with a projection along the azimuthal direction). Such fields are called optical vortices and carry OAM \cite{allen:pra92}. Axis-symmetrical plasmonic structures are known to convert SAM from the incoming light to OAM of the excited surface plasmons via optical spin-orbit interaction \cite{bliokh:sci15,gorodetski:prl09}. Optical spin-orbit interaction in plasmonic structures may thereby contribute to the generation of an optomagnetism in non magnetic metals via the magnetic electromotive force. The so-generated DC current density is then the result of an optical drag effect \cite{noginova:prb11,noginova:njp13} of circular symmetry.

\begin{figure}[hbt!]
\centering\includegraphics[width=1\linewidth]{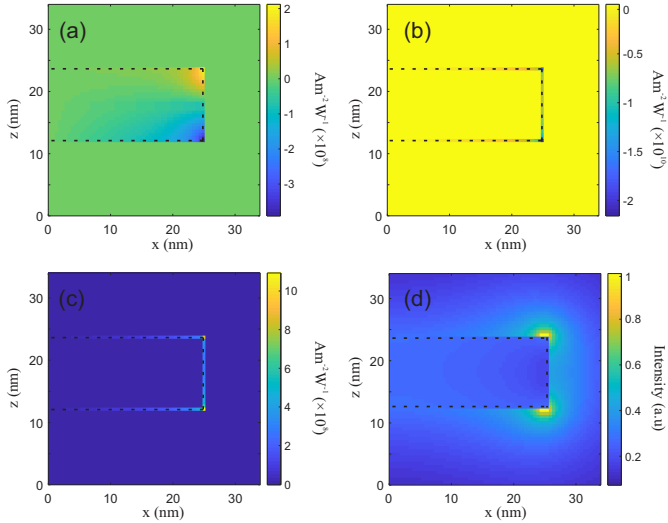}
\caption{(a-c) Distributions in a (x0z)-plane of the DC current densities induced by (a) the magnetic, (b) quadratic and (c) convective electromotive forces.(d) Electric field intensity in the (x0z)-plane of the plasmonic resonant mode of our nanoparticle.}\label{fig:current}
\end{figure}

Fig. \ref{fig:current}(a)-(c) represents the distributions of DC current in a cross-section of the nanodisk induced by the magnetic, quadratic and convective forces, respectively. Positive/negative values in the figures denote right/left-handed current densities, respectively.  Calculations are realized with a conductivity of silver at the zero frequency $\gamma_0=6.3 \cdot 10^7~S . m^{-1}$ \cite{lide:book}, an effective mass $m_e$ of the electron of 1.1 $\times 9.1091 \cdot 10^{-31}$kg \cite{johnson:prb72} and a density $n_e$ of free electrons of $5.86 \cdot 10^{28}$ \cite{lide:book}. The currents are expressed  in power unit of the incoming beam. In our case, a power of 1 Watt corresponds to a maximum intensity of $6.86 \cdot 10^{10} ~ W . cm^{-2}$ in the incident gaussian beam. Being proportionnal to the divergence of the polarization $\mathbf{P}$, both the quadratic and convective forces are mainly bound to the particle interfaces. As the amplitude of the magnetic force is negligible regarding the two other's, our model suggests that the magnetism induced optically via the free electrons gas is mainly a surface effect. The quadratic and convective forces have maximum value at the particle corners, a phenomenon imputed to the confinement and enhancement ability of plasmonic electric fields at metal discontinuities \cite{pile:apl05}, as shown in Fig. \ref{fig:current}(d). 

\begin{figure}[hbt!]
\centering
\includegraphics[width=1\linewidth]{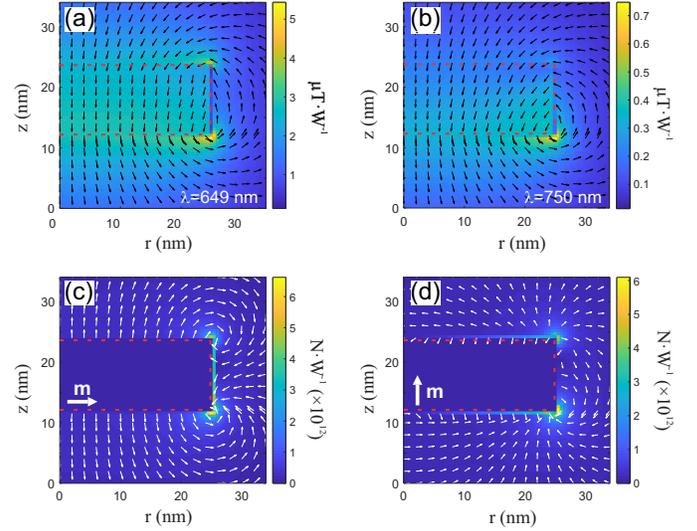}
\caption{(a) and (b) Distributions of the optically-induced static magnetic field in a longitudinal cross-section (r0z) of the plasmonic nanostructure, at wavelengths of 649 nm and 750 nm, respectively. (c,d) Optically-induced magnetic force in the (x0z)-plane exerted on a point-like particle of magnetic moment $\mathbf{m}$ of amplitude 1 oriented (c) along (0x) and (d) along (Oz).}\label{fig:Bfield}
\end{figure}

The key-role of metal corners in the confinement and enhancement of static magnetic fields is shown in Fig. \ref{fig:Bfield} (a). The static magnetic fields peaks at 5.43 $\mu$T/W right at the corners where it undergoes gradients much sharper than the particle sizes. For an incident electric intensity peaking at $10^{15}$ W.m$^{-2}$, the maximum of the static magnetic field would reach 79.2 mT, which is consistent with preceding theoretical works based on the inverse Faraday effect in plasmonic nanoparticles \cite{nadarajah:ox17}. From Eq. \ref{eq:momentangmag}, the induced magnetic moment is oriented along (0z) and its amplitude reaches $1.56 \cdot 10^{-22}$J.T$^{-1}$. The central role of the surface plasmons in the optomagnetic effect is evidenced by comparing Figs. \ref{fig:Bfield}(a) and (b) acquired with the plasmonic nanostructure on and off resonance. On-resonance (Fig. \ref{fig:Bfield}(a)), the nanostructure generates a static magnetic field that almost is one order of magnitude higher.

Figs. \ref{fig:Bfield} (d) and (c) plot in the (x0z)-plane the magnetic force exerted onto magnetic dipoles oriented along (0x) and (0z)  respectively. The amplitude of the magnetic dipole is 1. The maximum amplitude of the magnetic force reaches $6.61 \cdot 10^{12} ~ N.W^{-1}$ at the corners of the particle and it is sharply confined on the metal surface.

On the basis of a classical hydrodynamic model of the free electron gas of a metal, we investigate the ability of surface plasmons to generate a tiny DC current loop upon optical excitation with circular polarization. Our model is valid in spectral domains where the optical properties of metals are mainly governed by their free electrons. The optically-induced current density is found to be essentially confined at the metal surface and maximum at the particle's corners. The resulting static magnetic field and magnetic force are then deduced, leading to the concept of an optomagnet. Remarkably, the static magnetic field and force are shown to be sharply concentrated at the corners of a plasmonic nanodisk, owing to the unique ability of surface plasmons to be confined and enhanced at metal discontinuities. The design of axis-symmetrical nanoparticles with sharper corners should therefore notably enhance static magnetic fields. Surfaces plasmons at metal discontinuities thus provide new opportunities to generate and control static magnetic fields on the nanoscale and at optical frequencies, impacting a broad field of applications ranging from magneto-optics to spin waves. 






\end{document}